\documentclass[10pt,leqno]{amsart}
\usepackage{graphicx}
\usepackage[version=3]{mhchem}
\usepackage{indentfirst,csquotes}
\usepackage{geometry}

\usepackage{amssymb,amsthm,amsmath}
\usepackage{xcolor,paralist,hyperref,titlesec,fancyhdr,etoolbox}

\titleformat{\section}[display]{\normalfont\huge\bfseries\centering}{\centering\chaptertitlename\thechapter}{10pt}{\Large}
\titlespacing*{\section}{0pt}{0ex}{0ex}

\hypersetup{ colorlinks=true, linkcolor=black, filecolor=black, urlcolor=black }

\usepackage{lipsum}
\title{Seasonal variation in nighttime NO radiative cooling as observed by TIMED/SABER in lower thermosphere during solar maximum and solar minimum}
\author{Alok K Ranjan, MV Sunil Krishna$^\dagger$, Akash Kumar, Dayakrishna Nailwal, Sumanta Sarkhel}
\date{%
    $^1$Department of Physics, Indian Institute of Technology Roorkee, Roorkee-247 667, Uttarakhand %
    \today
}

\begin{document}

\begin{abstract}

Both composition and temperature play a crucial role in determining the NO radiative cooling in lower thermosphere \textcolor{black}{as observed by} TIMED/SABER. In this work, we present a detailed investigation of seasonal variation in thermospheric NO radiative cooling. We have carried forward the investigation of \cite{li2018} regarding the variations in local nighttime peak NO radiative cooling and its altitude during solar maximum and solar minimum conditions. \textcolor{black}{By analyzing latitudinal changes over quiet times for each month in year 2018, it is evident that both the investigative parameters exhibit summer-winter variability.} The qualitative contribution of different species (i.e., NO, and O), and temperatures in determining the vertical profile of NO radiative cooling for different latitudes is investigated by utilizing the NRLMSISE-00 estimated parameters, and SNOE observed NO density. The temperature, NO density, meridional wind, and associated compositional variations due to asymmetrical solar heating in both the hemispheres during solar minimum conditions seem to be the dominating factor in controlling the NO radiative cooling during different seasons. The altitudes at which maximum cooling by NO occurs exhibits an inverse correlation with the amount of radiative cooling. The region of enhanced NO densities (polar and summer hemispheric low-mid latitude regions) have larger NO radiative cooling with lower peak altitudes in comparison to other regions (equatorial to winter hemispheric low-mid latitude regions), where NO radiative cooling is low with higher peak altitude values.
\end{abstract}
\maketitle

\section*{Introduction}
Nitric oxide (NO) present in the lower to mid-thermosphere acts as a natural thermostat. It partially controls the outgoing energy of this region by converting its kinetic energy into radiative emission (5.3 $\mu$ m) through multiple collisions, thereby cooling the lower thermosphere \cite{kokart1980}. This process of NO radiative cooling is dominant during the solar maximum. Frequent geomagnetic storms during solar maximum result in enhanced energy input into the lower thermosphere in the form of Joule heating, and energetic particle precipitations in polar regions. The subsequent upwelling of N$_2$-rich air and their interaction with energetic precipitating electrons in polar regions during the energy deposition results in the dissociation of N$_2$-molecules into N($^2$D) and N($^4$S). These excited Nitrogen atoms \textcolor{black}{over polar latitudes} then interact with Oxygen molecules to form NO (Table-1; R$_1$ \& R$_2$), which can also quench back to N$_2$ molecules via temperature-based interaction with N($^4$S) (R$_3$ \& R$_4$). However, in the equatorial to mid- latitude regions the NO density is controlled by solar soft X-rays (2-10 nm) flux, and the meridional winds \cite{barth2003,barth2009}. The enhanced production of NO in the lower thermosphere during geomagnetic storms is also followed by its excitation in vibration-rotational band (\textcolor{black}{NO ($\textit{v} = 0$) to NO ($\textit{v} = 1$)}) through the temperature based collision rate with atomic oxygen (\textcolor{black}{R$_6$}), which then comes back to its ground state (\textcolor{black}{NO ($\textit{v} = 1$) to NO ($\textit{v} = 0$)}) by emitting a radiation of  5.3 $\mu$m. However, the atomic oxygen density can also cause the quenching of NO ($\textit{v} = 1$) (R$_5$), and hence the resultant of all these processes (R$_1$ to R$_6$) finally determines the amount of NO radiative cooling in lower thermosphere. 

There has been various studies on the behaviour of NO radiative cooling in the mesosphere and lower thermosphere during both extreme space weather conditions and forcing from below (atmospheric tides and sudden stratospheric warming events) \cite{mlynzack2003,mlynzack2005,mlynzack2010,ober2013,knipp2017,li2019,bag2021,ranjan2023a}.  The TIMED/SABER (Thermosphere-Ionosphere-Mesosphere Energetics and Dynamics/Sounding of the Atmosphere using Broadband Emission Radiometry) observed NO daily radiative power from thermosphere to space can reach upto hundreds of giga watts during geomagnetic storms with the enhancement of several tens of giga watts \textcolor{black}{compared} to quiet time values. The quiet time NO daily mean radiative power during solar minimum conditions can alone be of the order of tens of giga watts, and it reaches upto hundreds of giga watts during solar maximum \cite{lu2010}. This change in the outgoing radiative energy indicates the importance of NO infrared radiative emissions in controlling the energy budget of lower thermosphere during solar minimum and solar maximum.

As mentioned earlier, the total amount and pattern of NO radiative cooling are controlled by both composition (O$_2$, O, and NO) and temperature, which vary significantly through different seasons and at different latitude regions. The seasonal variation in outgoing energy loss of lower thermosphere in form of NO infrared radiative cooling will also depend on total incoming energy in different seasons. Since, the summer hemisphere gains more incoming energy in comparison to the winter hemisphere, we expect more NO infrared emissions from summer hemisphere. Due to the temperature gradient caused by the asymmetrical solar heating between the hemispheres, meridional winds from summer to winter in the lower thermosphere also result in a higher concentration of lighter species (oxygen atoms) in the winter hemisphere. Heavier species (O$_2$, and N$_2$), however, are concentrated in the summer hemisphere (\cite{Brasseur2005}). As a result, it would also affect the column O/N$_2$ ratios at different altitudes and latitudes through different seasons. 

Study of thermospheric NO radiative cooling patterns and associated thermospheric density response can help in predicting satellite drag in earth's upper atmosphere during different solar and geophysical conditions \cite{knipp2013,li2019}. The changing heat budget of atmosphere may also modulate other atmospheric and ionospheric parameters that indirectly affect human life at the surface. Some examples are, Lower Earth Orbit (LEO) satellite drags (which are particularly used for geophysical imaging), GPS navigation, exposure of thermospheric particles near ISS (International Space stations), etc. All these aspects need to be considered for a better preparation and to mitigate the potential threats due to space weather influence in future. In this work, we investigate the variations in peak NO radiative cooling, and its altitude in the local nighttime through different seasons in different latitude regions. This work goes further to establish the responsible thermospheric species or variable for the observed variability during the solar minimum conditions. The data sources and data processing used in this study are presented in Section 2. Section 3 represent the results and discussion of our investigation, and the conclusions are presented in Section 4.

\section*{Data analysis}
\label{sec2}
In December 2001, the TIMED satellite was launched into a 625 km near-circular orbit with a 74$^\circ$ inclination. Every 60 days, orbit precession allows for 24-hour local time coverage, and TIMED conducts a yaw movement to keep SABER from pointing to the Sun, resulting in only few SABER observations around local noon \cite{marty1997,esplin2023}. Depending on the yaw cycle, the recorded geographic latitude ranges from 83$^\circ$ in one hemisphere to 55$^\circ$ in the other. The SABER instrument is a 10-channel infrared radiometer that sweeps the Earth's limb from the 400-km tangent height to the surface, taking an infrared radiance sample every 0.4 km. The instrument's spectral coverage ranges from 1.27 to 15.4 $\mu m$, including NO at 5.3 $\mu m$. Since most \textcolor{black}{of} the observational data from SABER (Level2A in-band NO infrared emissions) is in nighttime, we have only utilized its nighttime datasets \textcolor{black}{(between 18 Solar Local Time (SLT) to 6 SLT through midnight), the TIMED/SABER observed SLT datasets have been considered for this classification.} 

\textcolor{black}{To investigate the seasonal variation in peak NO radiative cooling and its corresponding altitude in absence of geomagnetic activity, first three IQDs (q1, q2, q3) are considered for each months during solar minimum (year 2018). The daily averages between q1, q2, and q3 with 20$^\circ$ latitudinal resolution are implemented to establish the seasonal variation for both the parameters.} Empirical atmospheric model NRLMSISE-00 (Naval Research Laboratory Mass Spectrometer and Incoherent Scatter Radar Exosphere) output datasets are used in different latitude and altitude regions with 31 days running average also only for the nightside. SNOE (Student Nitric Oxide Explorer) observational datasets are also used to understand the variations of NO density throughout the seasons in different latitude regions.

\section*{Results and Discussion}
\label{sec3}
\begin{figure*}
\centering
\includegraphics[width=7.3in,height=6.0in]{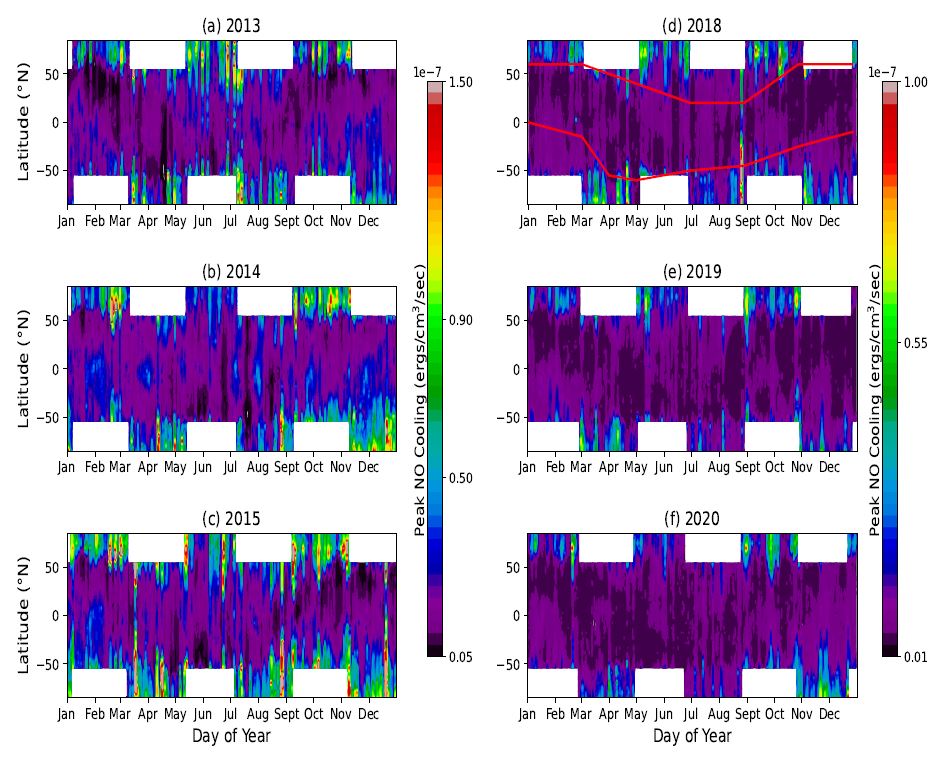}
\caption{Latitudinal \textit{or} seasonal variation in peak NO radiative cooling in lower thermosphere throughout solar maximum ((a) 2013, (b) 2014, (c) 2015) and solar minimum ((d) 2018, (e) 2019, (f) 2020) of solar cycle 24.}
\label{fig:figure1}
\end{figure*}

\begin{figure*}[h]
\includegraphics[width=7.3in,height=6.0in]{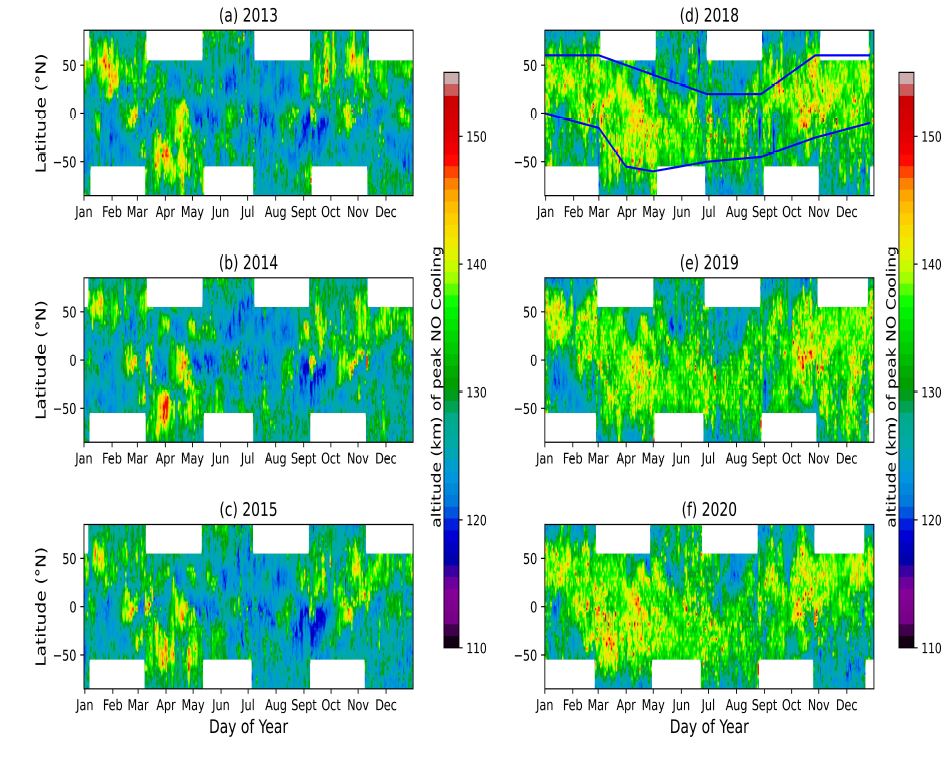}
\centering
\caption{Latitudinal \textit{or} seasonal variation in the altitude of peak NO radiative cooling throughout solar maximum ((a) 2013, (b) 2014, (c) 2015) and solar minimum ((d) 2018, (e) 2019, (f) 2020) of solar cycle 24. \textcolor{black}{The blue lines are drawn to mostly represent the regions which have larger peak altitudes at a particular DOY compared to other regions.}}
\label{fig:figure2}
\end{figure*}

As mentioned earlier, in this study we investigate the temperature and compositional (O, and NO) contribution to the TIMED/SABER observed nighttime NO peak radiative cooling and its altitude in different latitude regions. \cite{li2018} presented the seasonal variations in peak NO radiative cooling at different local time and latitudes and compared it with TIE-GCM (Thermosphere-Ionosphere-Electrodynamics General Circulation Model) simulated results. However, the contribution of atmospheric parameters (such as  perturbed dynamics, composition, and temperature) accounting to such variation is still an open question. Figure \ref{fig:figure1} and Figure \ref{fig:figure2} show the nighttime peak NO radiative cooling and its corresponding altitude, respectively. The variation in these parameters is shown at different latitude regions during both solar maximum ((a) 2013, (b) 2014, and (c) 2015) and solar minimum ((d) 2018, (e) 2019, and (f) 2020) conditions. \textcolor{black}{The 27-day averaged F10.7 index during 24 solar cycle (figure not shown here) also shows that year 2013, 2014, and 2015 lie within the solar maximum, while year 2018, 2019, and 2020 lies within the solar minimum period.} It is to be noted that only the peak emission rate of a given profile is taken for representation in these figures. It can be seen in Figure \ref{fig:figure1} that during solar maximum and solar minimum conditions most of the peak NO radiative cooling occurs in high latitude regions of both the hemispheres, with its larger equator-ward penetration from summer hemisphere. The colorbar limit in the figure during both the conditions are different, which also suggests that the peak cooling is also larger during the solar maximum in comparison to solar minimum, as expected.

\textcolor{black}{It is to be noted that during solar minimum conditions, high speed solar wind (HSSW) and co-rotating interaction region (CIR) can caused minor storms. The geomagnetic A$_p$ index throughout the year 2018 (figure not shown here) suggests that there are at least 6 minor storms (18-19 March, 20 April, 5-6 May, 13 September, 7 October, and 5 November) and one major geomagnetic storm (26 August) present during this period. As mentioned earlier, presence of these storms can be seen in form of enhanced peak NO cooling and its penetration towards the equator from both polar regions. In particular, it can be seen in Figure \ref{fig:figure1} (d) that peak NO cooling is enhanced in about every latitude region during the major geomagnetic storm of 26 August, which was driven by aggregation of a weak coronal mass ejection (CME) transients and CIR/HSSs}

It is interesting to note in Figure \ref{fig:figure1} and Figure \ref{fig:figure2} that the both peak NO radiative cooling and its altitude show a seasonal variation, particularly during solar minimum conditions. During solar minimum conditions, the NO peak emission rates in low to mid-latitude regions are larger with lower peak altitudes in the summer hemisphere in comparison to winter hemisphere. However, during solar maximum conditions, the seasonal variation in the altitude of peak NO radiative cooling get disrupted as can be seen in Figure \ref{fig:figure2} (a, b, and c). The strongly perturbed meridional winds during solar maximum conditions can play a major role in determining the global NO radiative cooling. It has been reported in the past that meridional winds can control the global distribution \textcolor{black}{of} both NO density, and NO radiative cooling \cite{barth2010,li2019,bag2021}. As a result, we can expect a perturbation in seasonal variation of peak NO radiative cooling and its respective altitude during the solar maximum. 

Figure \ref{fig:figure2} shows that during solar minimum, altitude of peak NO radiative cooling is mostly restricted to the northern hemisphere during the months of October through February with an average of about 140 km. Meanwhile,  during the months of April through August the peak is located in the southern hemisphere. The latitude regions \textcolor{black}{manually} bounded by the blue lines throughout the year 2018 shows this variation. The same boundary line is indicated with a red line in Figure \ref{fig:figure1} (d), which shows that, for almost all days, the peak NO radiative cooling has a minima in this bounded red lines regions. 

\begin{figure*}[h]
\includegraphics[width=6.8in,height=6.3in]{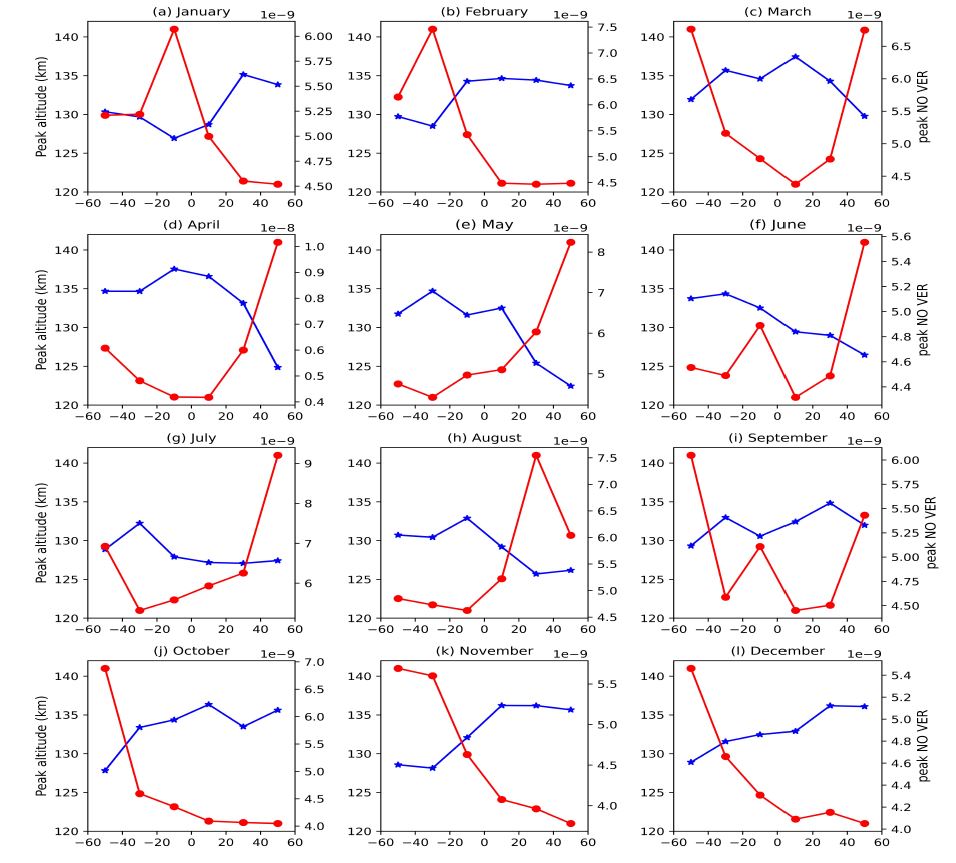}
\centering
\caption{(a-l) Latitudinal \textit{or} seasonal variation in the altitude of peak NO radiative cooling throughout quiet days for year 2018. The blue lines represent the peak altitude (km) and red lines represent the corresponding peak NO radiative cooling (ergs/cm$^3$/sec) values.}
\label{fig:figure3}
\end{figure*}

\begin{figure*}[h]
\includegraphics[width=4in,height=4.0in]{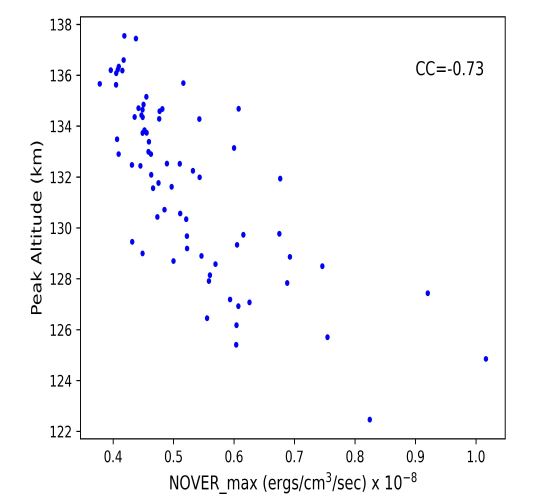}
\centering
\caption{Correlation of peak NO radiative cooling with its corresponding altitude for quiet days of year 2018.}
\label{fig:figure4}
\end{figure*}

\textcolor{black}{To understand the seasonal variation of peak altitude of NO radiative cooling (Figure \ref{fig:figure1} and Figure \ref{fig:figure2}) in the absence of geomagnetic activity, three international quiet days (IQDs) are considered in each month throughout year 2018. In Figure \ref{fig:figure3} (a-l), daily nighttime mean of both the investigative parameters are shown with 20$^\circ$ latitude resolution (40-60 $^\circ$, 20-40 $^\circ$, 0-20 $^\circ$). This figure establishes that, during northern summer months (May, June, July) northern hemisphere witnesses lower peak altitude in comparison to southern hemisphere. Similarly, during winter months (November, December, January, February) in northern hemisphere, it can be clearly seen that the southern hemisphere has lower peak altitude in comparison to northern hemisphere. It is also to be noted from the figure that, the variation in NO radiative cooling is exactly opposite to what is seen in peak altitude. A negative correlation coefficient of -0.73 is found (Figure \ref{fig:figure4}) by considering the same datasets utilized in Figure \ref{fig:figure3}. That is, in summer hemisphere peak NO radiative cooling profile has a larger value with lower peak altitude in comparison to winter hemisphere.} This aspect is further elaborated by utilizing the NRLMSISE-00 datasets in upcoming paragraphs. 

\begin{figure*}[hbt!]
\includegraphics[width=5in,height=7.5in]{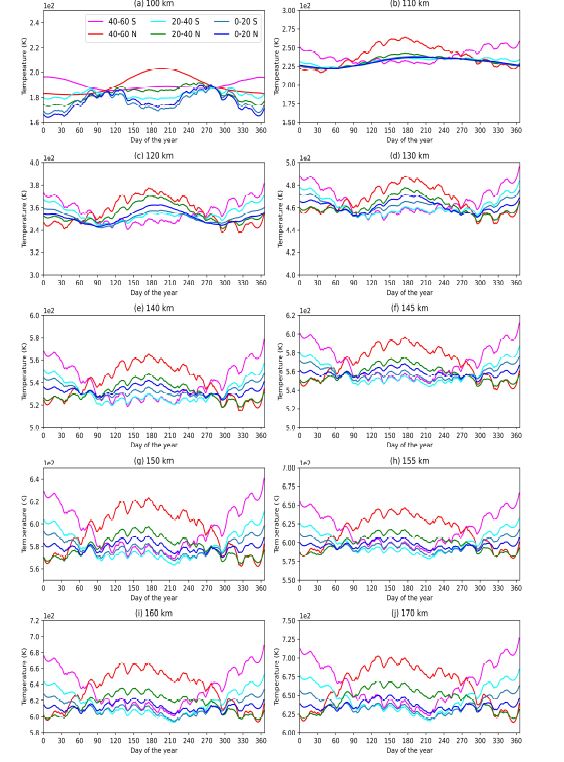}
\centering
\caption{Seasonal variation in nighttime temperature at different altitudes and in different latitude zones throughout the year 2018 as estimated by NRLMSISE-00.}
\label{fig:figure5}
\end{figure*}

Figure \ref{fig:figure5} (a-j) shows the \textcolor{black}{NRLMSISE-00} estimated 31 day running average variation in temperature at various latitude sectors for a given altitude.  A clear seasonal variation in temperature at all the altitudes, and latitudes can be seen between 110 to 170 km. This can be mainly due to the asymmetrical solar heating between both the hemispheres. From April to August, the temperature in northern hemisphere is larger in comparison to the southern hemisphere with its larger values towards the poles. The same is true for southern hemisphere during the months between October to February. A wave like pattern in temperature can be seen in almost every latitude and altitude region in lower thermosphere, which could be due to the effect of solar EUV \textit{or} captured by the F10.7 index variations, which also follows the 27-day solar rotation period. As the active chemical reactions (Table 1) responsible for 5.3 $\mu m$ emission by NO depend on temperature, it would be interesting to see the effects of seasonally \& latitudinally varying temperature on the peak NO radiative cooling emission and its corresponding altitude.

\begin{figure*}[hbt!]
\includegraphics[width=5.8in,height=7.5in]{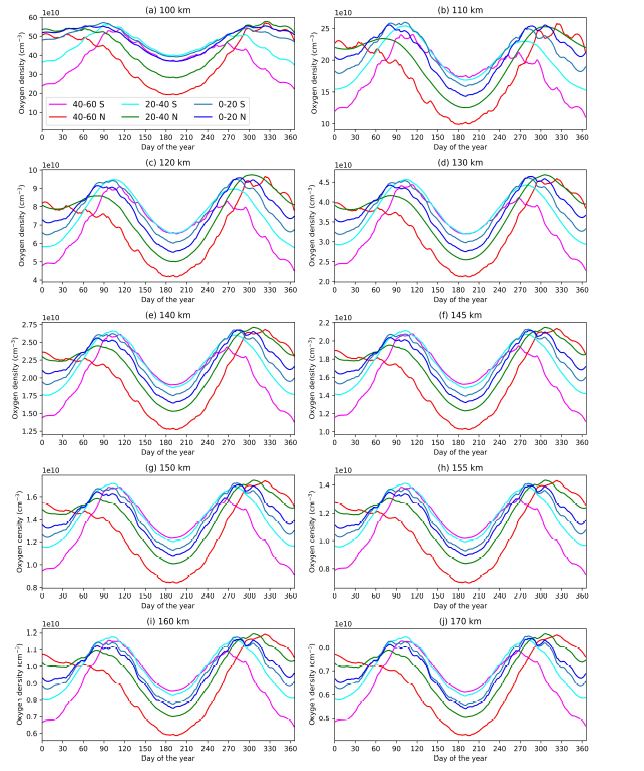}
\centering
\caption{Seasonal variation in nighttime oxygen density at different altitudes  and in different latitude zones throughout the year 2018 as estimated by NRLMSISE-00.}
\label{fig:figure6}
\end{figure*}

The collisional excitation rate of NO (\textit{v} = 0) to NO (\textit{v} = 1) strongly depends on temperature along with the concentration of atomic oxygen (collisional excitation rate C01 (sec$^{-1}$) = [O] $\times$  $4.2\times10^{-11} \exp({-2700/\textrm{T}})$) (\cite{hwang2003,mlynzack2021}), which makes atomic oxygen an important constituent in controlling the NO radiative cooling emissions. Figure \ref{fig:figure6} (a-j) shows the seasonal variation in atomic oxygen density for different altitudes and latitudes same as Figure \ref{fig:figure5} (a-j). It can be seen in the figure that, the variation in atomic oxygen is in opposite sense of temperature, i.e. (1) its density is decreasing with altitude and (2) its density is larger in northern hemisphere during the months of October to February and in the southern hemisphere during the months of April to August. These variations are expected due to expansion of the atmosphere during larger temperature, and higher susceptibility of oxygen atoms to the dynamical transport by winds induced by asymmetric solar heating. 

\begin{figure*}[hbt!]
\includegraphics[width=5.8in,height=3.5in]{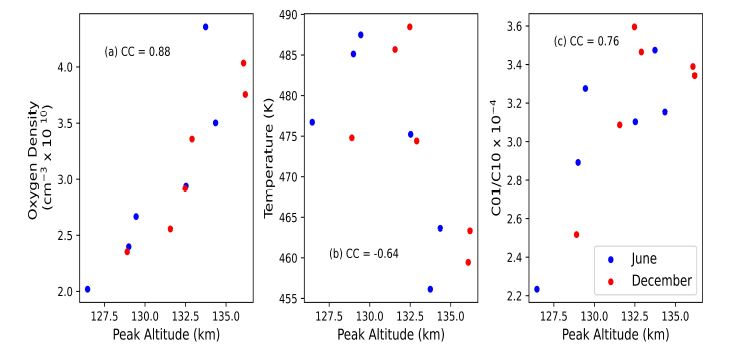}
\centering
\caption{Correlation of peak altitude of NO radiative cooling with (a) atomic oxygen, (b) temperature, and (c) C01/C10 for June and December months throughout the quiet days of year 2018.}
\label{fig:figure7}
\end{figure*}

An engrossing correlation between the atomic oxygen density and the altitude of peak NO radiative cooling (Figure \ref{fig:figure2} (d)) in different latitude regions is also noticeable. In January, February, November, and December, it can be seen in the figures that both oxygen density, and altitude of peak NO radiative cooling are highest in the northern hemisphere. From mid March to April the oxygen density reaches a peak with comparable concentration in every latitude regions except for the northern polar region. At the same time, the observed altitude of peak cooling emissions are also elevated in almost all the latitude regions between 40$^\circ$N to 60$^\circ$S. During May to August the observed enhanced peak altitude of NO cooling is limited to 10$^\circ$N to 60$^\circ$S, where the atomic oxygen density (Figure \ref{fig:figure6}) is also higher in southern hemisphere in comparison to the northern. \textcolor{black}{A positive correlation of 0.88 is also found between atomic oxygen density and altitude of peak NO radiative cooling, while a negative correlation of -0.64 is found between temperature and altitude of peak NO radiative cooling (Figure \ref{fig:figure7}(a \& b)) by considering the IQDs datasets for June and December months of year 2018.}

\begin{figure*}[hbt!]
\includegraphics[width=5.2in,height=7.5in]{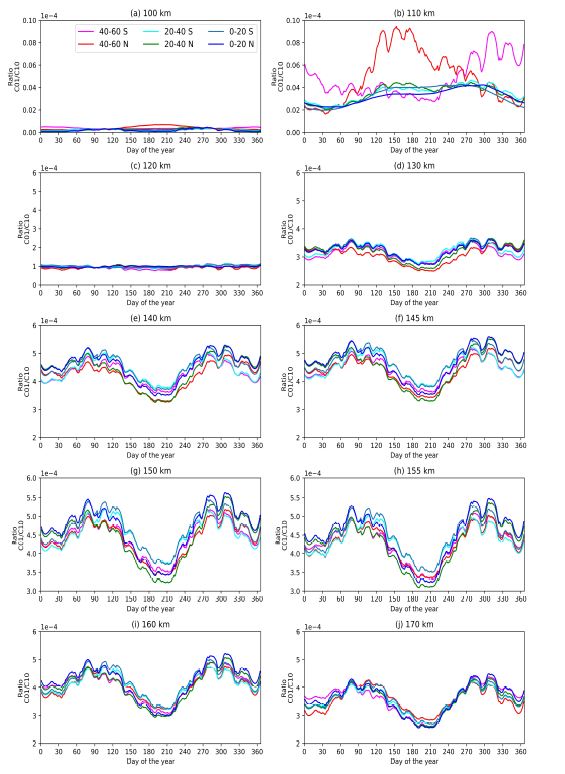}
\centering
\caption{Seasonal variation in nighttime NO (\textit{v} = 1) population ratio (C01/C10) at different altitudes  and in different latitude zones throughout the year 2018 as estimated by NRLMSISE-00.}
\label{fig:figure8}
\end{figure*}

The population ratio of excitation (\textcolor{black}{NO ($\textit{v} = 0$) to NO ($\textit{v} = 1$)}) and quenching (\textcolor{black}{NO ($\textit{v} = 1$) to NO ($\textit{v} = 0$)}) rate of NO with atomic oxygen is shown in Figure \ref{fig:figure8} (a-j) in the same format as Figure 5 \& 6 (population ratio; C01/C10 = 4.2$\times$10$^{-11}$$[O]$$\exp({-2700/\textrm{T}})$/(A$_{5.3\mu m}$ + 4.2$\times$10$^{-11}$$[O]$); where A$_{5.3\mu m}$ is the Einstein coefficient for spontaneous emission (12.32 sec$^{-1}$)). It is evident from the figure that the population ratio in every latitude regions is increasing with height and reaching a maximum around 145-150 km, after which it again starts decreasing very slowly. There is no clear seasonal variation till the altitude of 130 km, after that a clear enhancement in population ratio from January to March and a decrement from April to July can be seen. During summer months in northern hemisphere  (May to August), the population ratio is larger in southern hemisphere dominating near low to mid latitude regions (0-40 $^\circ$S) and reaches a minimum near mid-latitudinal northern hemisphere. Similarly, during the winter months in northern hemisphere (November to February) the ratio is larger near 0-40 $^\circ$N and reaches a minimum near mid-latitudinal southern hemisphere. During the months of late March, April, late September, and October, the population ratio reaches its highest in low latitude regions (0-20$^\circ$N, and 0-20$^\circ$S) followed by the mid (20-40$^\circ$N, and 20-40$^\circ$S) regions of both the hemisphere. \textcolor{black}{The reason behind this particular enhancement of C01/C10 could be due to the increase in atomic oxygem density in winter hemisphere, and near equinoxes.  The enhancement near equinoxes are supposed to be due to the semiannual variation in thermospheric density and composition (\cite{fuller1998}). It is proposed that density scale height is higher during equinoxes than solstices due to less mixing caused by lower large-scale inter-hemisphere circulation, resulting in semiannual variation in neutral density. The effect of this semiannual variation is also present in TIMED/SABER observed nighttime NO radiative cooling profiles in form of its large peak altitudes (Figure \ref{fig:figure2}) during the same periods (March, April, September, and October).} \textcolor{black}{Ratio C01/C10 also have a positive correlation (0.76; Figure \ref{fig:figure7} (c)) with altitude of peak NO radiative cooling during IQDs of June and December months of year 2018. This indicates that regions with larger C01/C10 (winter hemisphere) have larger peak altitude of NO radiative cooling.}

\begin{figure*}[hbt!]
\includegraphics[width=6.2in,height=6.50in]{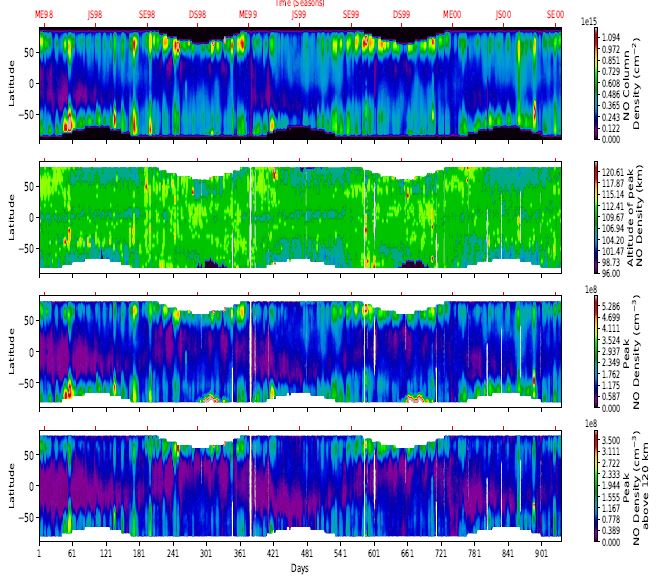}
\centering
\caption{Latitudinal \textit{or} seasonal variation in (a) NO column density, (b) altitude of peak NO density, (c) peak NO density, and (d) peak NO density above 120 km as observed by SNOE. \textcolor{black}{This plot corresponds to 1998-2000 period.}}
\label{fig:figure9}
\end{figure*}

In addition to the temperature and oxygen abundance, NO density also plays an important role in determining the NO radiative cooling profile (\cite{bharti2018,ranjan2023a}). \textcolor{black}{Since there is no continuous observational data of thermospheric NO density during the 24 solar cycle, SNOE observed datasets have been utilized to observe the seasonal variation in NO density in every latitude range.} Figure \ref{fig:figure9} \textcolor{black}{represents} the SNOE observed NO density profile for the 935-day period (from 11 March 1998 to 30 September 2000) (as in \cite{barth2003}). At the start of the 935-day period, the SNOE observations were taken at a mean local time of 10:17 AM, which shifted to 11:11 AM at the conclusion of the period. In the polar regions, NO is created by energetic electron precipitation in nightside and detected during the day when the Earth's rotation has shifted the location of electron precipitation into sunlight. In the mid to low latitude regions, only solar radiation (soft X-rays, EUV), and meridional winds are responsible for the variability of NO density. Consequently, it can also be seen in both the figures (Figure \ref{fig:figure9}; top panel and second panel from below) that equatorial to mid latitude NO density is enhanced in summer hemisphere, which also sometimes seems to penetrate into winter hemisphere possibly due to strong summer to winter meridional winds. Figure \ref{fig:figure9} (second panel from above) represents the altitude of peak NO density observed by SNOE. It is evident in the figure that there is no clear asymmetry in the peak altitude between both the hemispheres, which is about 105 km in high latitude and about 110 km in mid latitude regions during solstices. 



The enhanced NO density profile competes with C01/C10 population ratio profile to determine the altitude of peak NO radiative cooling emissions. It is evident in Figure \ref{fig:figure2}(d) that, in most of the regions peak altitude NO cooling is above 120 km. Figure \ref{fig:figure9} (lower panel) represents the peak NO density above 120 km, which also shows the same seasonal variation as Figure \ref{fig:figure9} (second panel from below).


Both the peak NO density (Figure \ref{fig:figure9}; lower two panel) and peak NO cooling emissions (Figure \ref{fig:figure1}(d)) have their minima in low-mid latitude of winter hemisphere. However, the peak NO cooling altitude in the same region are highest above 135 km (Figure \ref{fig:figure2} (d)) compared to other regions. Hence, it might be concluded that, in these particular regions due to the lower NO density, the peak altitude of NO cooling is controlled mostly by the temperature and atomic oxygen in the form of population ratio of NO (C01/C10), which approximately peaks at about 145 km. At the same time, for the low-mid latitude of summer hemisphere, the contribution of enhanced NO densities particularly at altitude about 120-130 km dominates over the population ratio (C01/C10) (i.e., temperature \& oxygen) in determining the altitude of peak NO cooling emission rates, which is at about 125-130 km.

\begin{figure*}[hbt!]
\includegraphics[width=5.2in,height=4.20in]{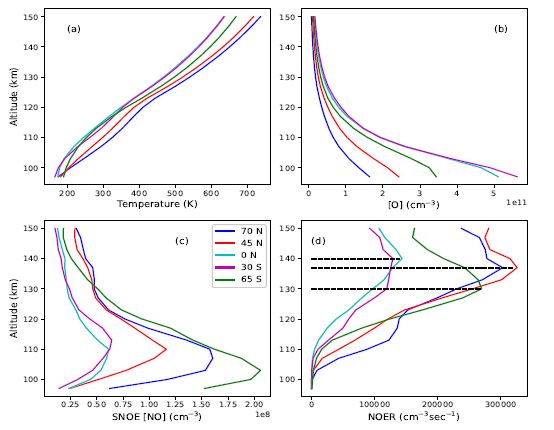}
\centering
\caption{Altitudinal variation in NRLMSISE-00 estimated (a) temperature, (b) O density, (c) SNOE observed NO density, and (d) NRLMSISE-00 \& SNOE estimated NOER at different latitudes and 0 E longitude at 10 UT on 21 June 1998.}
\label{fig:figure10}
\end{figure*}

To establish the seasonal variation in altitude of peak NO radiative cooling using SNOE datasets, the latitudinal variation in SNOE estimated peak NO radiative cooling and its corresponding altitude is also represented in Figure \ref{fig:figure10} (a-d) during summer solstice (21 June) of 1998. NRLMSISE-00 estimated (a) temperature and (b) oxygen density altitude profile is also shown at about 10 LST with (c) SNOE's observational path of NO density at different latitudes, which are also used to estimate the (d) NO emission rate (NOER = C01/C10 $\times$ [NO]). It can be seen from the figure that , in polar and mid latitude region of summer (northern) hemisphere, the altitude of peak NOER is lower (136 km) in comparison to equatorial to mid-latitude region of winter hemisphere (140 km) (Figure \ref{fig:figure10}(d)).  In the winter polar region, again due to \textcolor{black}{enhanced} NO density in lower thermosphere, the altitude of peak NO cooling becomes lower (130 km) with enhanced NO cooling (Figure \ref{fig:figure10}(c \& d)). \textcolor{black}{The SNOE estimated latitudinal/seasonal variation in the altitude of peak NOER are higher than observed by TIMED/SABER (125-130 km) (Figure \ref{fig:figure2}). However, their qualitative variation is parallel irrespective of their different time of observation. This suggest that NO density profile between 120-140 km is very crucial to determine the peak NO radiative cooling and its altitude in every season.}

\section*{Conclusions}
\label{sec4}
In this study, we have investigated the contribution of composition, and temperature in the TIMED/SABER observed peak NO cooling emission and its altitude. \textcolor{black}{By comparing low- to mid- latitudinal variations during the first three IQDs for each months of year 2018, it was clearly seen that there is a strong summer-winter variability in both peak NO radiative cooling emission and its corresponding altitude.} The region of enhanced NO density (polar, and low to mid latitude of summer hemisphere) particularly between 120-140 km leads to larger NO radiative cooling with lower peak altitude (125-130 km). The equatorial to mid latitude of winter hemisphere have low NO density, due to which NO radiative cooling is low with its peak altitude at about 140 km. The hemispheric asymmetry of [NO] and [O] due to asymmetric solar heating between both the hemispheres seem to be the root cause of such variations during solar minimum conditions. Both the peak NO radiative cooling emissions, and its peak altitude are controlled by a combined effect of [O], [NO], \textcolor{black}{and} temperature variations. The contribution of [O], and temperature in form of population \textcolor{black}{ratio} of NO ($v$ = 1)/NO ($v$ = 0) (C01/C10) is dominant in the low to mid latitudes of winter hemisphere. This is very crucial in determining the vertical profile of NO radiative cooling in winter hemisphere. The three parameters ([NO],[O],T) responsible for the net radiaitve flux at 5.3 $\mu$m seems to vary as a function of latitude and altitude. During solar maximum conditions, these seasonal pattern could get disrupted due to storm induced intensified meridional winds.

\section*{Acknowledgments}
We thank TIMED/SABER, NRLMSISE-00 and SNOE science teams for providing the data ($http://saber\-.gats-Inc.com/browse\_data.php$), ($https://ccmc.gsfc.nasa.gov/modelweb/models/nrlmsise00.php$), ($https://lasp\-.colorado.edu/snoe/data/$) used in this study. One of the author AKR thanks Ministry of Human Resource Development, New Delhi for the financial support in the form of graduate assistantship.

\pagebreak


\pagebreak

\begin{table*}
\centering
\caption{NO Reaction Rate Coefficients}
\begin{tabular}{ p{6.5cm} p{6.0cm} p{3.0cm} }
\hline
Reaction & Rate Coefficient & Reference \\ 
\hline
\ce{(R1) N($^2$D) + O$_2$ $\rightarrow$ NO + O} & $6.2\times10^{-12}(\textrm{T}/300)$ &\cite{duff2003}\\ 
\ce{(R2) N($^4$S) + O$_2$ $\rightarrow$ NO + O} & $1.5\times10^{-11}\exp({-3600/\textrm{T}})$ &\cite{siskind1992}\\
\ce{(R3) N($^4$S) + NO $\rightarrow$ N$_2$ + O} & \textrm{T}$>$400 K:$3.25\times10^{-11}$ &\cite{swaminathan1998}\\
\ce{(R4) N($^4$S) + NO $\rightarrow$ N$_2$ + O} & T$<$400 K:$2.2\times10^{-11}\exp({160/\textrm{T}})$ &\cite{swaminathan1998}\\ 
\ce{(R5) NO($v$=1) + O $\rightarrow$[$\textit{k$_{10}$}$] NO($v$=0) + O} & $4.2\times10^{-11}$ &\cite{hwang2003}\\
\ce{(R6) NO($v$=0) + O $\rightarrow$[$\textit{k$_{01}$}$] NO($v$=1) + O} & $4.2\times10^{-11} \exp({-2700/\textrm{T}})$ &\cite{hwang2003,mlynzack2021}\\
\hline
\end{tabular}
\end{table*}

\begin{thebibliography}{250}
\bibitem{li2018}
Li, Z., Knipp, D., Wang, W., Sheng, C., Qian, L. and Flynn, S., \href{https://agupubs.onlinelibrary.wiley.com/doi/full/10.1029/2018JA025831}{A comparison study of NO cooling between TIMED/SABER measurements and TIEGCM simulations. Journal of Geophysical Research: Space Physics {\bf{123(10)}}, 2018.}
\bibitem{kokart1980}
Kockarts, G., \href{https://agupubs.onlinelibrary.wiley.com/doi/abs/10.1029/GL007i002p00137}{Nitric oxide cooling in the terrestrial thermosphere. Geophysical Research Letters, {\bf{7(2)}}, 137-140, 1980.}
\bibitem{barth2003}
Barth, C.A., Baker, D.N. and Bailey, S.M., \href{https://agupubs.onlinelibrary.wiley.com/doi/full/10.1029/2003GL018892}{Seasonal variation of auroral electron precipitation. Geophysical research letters {\bf{31(4)}}, 2003.}
\bibitem{barth2009}
Barth, C.A., Lu, G. and Roble, R.G., \href{https://agupubs.onlinelibrary.wiley.com/doi/full/10.1029/2008JA0137655}{Joule heating and nitric oxide in the thermosphere. Journal of Geophysical Research: Space Physics {\bf{114(A5)}}, 2009.}
\bibitem{mlynzack2003}
Mlynczak, M., Martin-Torres, F.J., Russell, J., Beaumont, K., Jacobson, S., Kozyra, J., Lopez-Puertas, M., Funke, B., Mertens, C., Gordley, L. and Picard, R., \href{https://agupubs.onlinelibrary.wiley.com/doi/full/10.1029/2003GL017693}{The natural thermostat of nitric oxide emission at 5.3 ?m in the thermosphere observed during the solar storms of April 2002. Geophysical Research Letters {\bf{30(21)}}, 2003.}
\bibitem{mlynzack2005}
Mlynczak, M.G., Martin-Torres, F.J., Crowley, G., Kratz, D.P., Funke, B., Lu, G., Lopez?Puertas, M., Russell III, J.M., Kozyra, J., Mertens, C. and Sharma, R., \href{https://agupubs.onlinelibrary.wiley.com/doi/abs/10.1029/2001JA001107}{Energy transport in the thermosphere during the solar storms of April 2002. Journal of Geophysical Research: Space Physics {\bf{110(A12)}}, 2005.}
\bibitem{mlynzack2010}
Mlynczak, M. G., Hunt, L. A., Thomas Marshall, B., Martin$-$Torres, F. J., Mertens, C. J., Russell III, J. M., ... \& Gordley, L. L., \href{https://agupubs.onlinelibrary.wiley.com/doi/full/10.1029/2009JA014713}{Observations of infrared radiative cooling in the thermosphere on daily to multiyear timescales from the TIMED/SABER instrument. Journal of Geophysical Research: Space Physics, {\bf{115(A3)}}, 2010.}
\bibitem{ober2013}
Oberheide, J., M. G. Mlynczak, C. N. Mosso, B. M. Schroeder, B. Funke, and Astrid Maute, \href{https://agupubs.onlinelibrary.wiley.com/doi/full/10.1002/2013JA019278}{Impact of tropospheric tides on the nitric oxide 5.3 $\mu$m infrared cooling of the low‐latitude thermosphere during solar minimum conditions. Journal of Geophysical Research: Space Physics, {\bf{118(11)}}, 2013.}
\bibitem{knipp2017}
Knipp, D.J., Pette, D.V., Kilcommons, L.M., Isaacs, T.L., Cruz, A.A., Mlynczak, M.G., Hunt, L.A. and Lin, C.Y., \href{https://agupubs.onlinelibrary.wiley.com/doi/full/10.1002/2016SW001567}{Thermospheric nitric oxide response to shock?led storms. Space Weather {\bf{15(2)}}, 2017.}
\bibitem{li2019}
Li, Z., Knipp, D. and Wang, W., \href{https://agupubs.onlinelibrary.wiley.com/doi/full/10.1029/2018JA026247}{Understanding the behaviors of thermospheric nitric oxide cooling during the 15 May 2005 geomagnetic storm. Journal of Geophysical Research: Space Physics {\bf{124(3)}}, 2019.}
\bibitem{bag2021}
Bag, T., \href{https://www.sciencedirect.com/science/article/pii/S0273117721003732}{Local-time, seasonal and solar cycle variation of Nitric Oxide radiative emission over Indian longitude sector. Advances in Space Research {\bf{68(6)}}, 2021.}
\bibitem{ranjan2023a}
Ranjan, A.K., Sunil Krishna, M.V., Kumar, A., Sarkhel, S., Chakrabarty, D. and Reeves, G.D., \href{https://agupubs.onlinelibrary.wiley.com/doi/full/10.1029/2023JA032028}{{\sl NO Radiative Cooling and Ionospheric Response to the HILDCAA Events Following Geomagnetic Storms. Journal of Geophysical Research: Space Physics, {\bf{128(12)}}, p.e2023JA032028, (2023)}}
\bibitem{lu2010}
Lu, G., Mlynczak, M.G., Hunt, L.A., Woods, T.N. and Roble, R.G., \href{https://agupubs.onlinelibrary.wiley.com/doi/full/10.1029/2009JA014662}{On the relationship of Joule heating and nitric oxide radiative cooling in the thermosphere. Journal of Geophysical Research: Space Physics {\bf{115(A5)}}, 2010.}
\bibitem{Brasseur2005}
Brasseur, G. P., \& Solomon, S., \href{https://link.springer.com/book/10.1007/1-4020-3824-0}{Aeronomy of the middle atmosphere: Chemistry and physics of the stratosphere and mesosphere ({\bf{Vol. 32}}). Springer Science \& Business Media, 2005.}
\bibitem{knipp2013}
Knipp, D., Kilcommons, L., Hunt, L., Mlynczak, M., Pilipenko, V., Bowman, B., Deng, Y. and Drake, K., \href{https://agupubs.onlinelibrary.wiley.com/doi/full/10.1002/grl.50197}{Thermospheric damping response to sheath-enhanced geospace storms. Geophysical Research Letters {\bf{40(7)}}, 2013.}
\bibitem{marty1997}
Mlynczak, M.G., \href{https://www.sciencedirect.com/science/article/pii/S0273117797007692}{Energetics of the mesosphere and lower thermosphere and the SABER experiment. Advances in Space Research {\bf{20(6)}}, pp.1177-1183, 1997.}
\bibitem{esplin2023}
Esplin, R., Mlynczak, M.G., Russell, J., Gordley, L. and SABER Team., \href{https://agupubs.onlinelibrary.wiley.com/doi/full/10.1029/2023EA002999}{Sounding of the Atmosphere using Broadband Emission Radiometry (SABER): Instrument and science measurement description. Earth and Space Science {\bf{10(9)}}, p.e2023EA002999, 2023.}
\bibitem{barth2010}
Barth, C.A.,  \href{https://agupubs.onlinelibrary.wiley.com/doi/full/10.1029/2010JA015565}{Joule heating and nitric oxide in the thermosphere, 2. Journal of Geophysical Research: Space Physics {\bf{115(A10)}}, 2010.}
\bibitem{hwang2003}
Hwang, E.S., Castle, K.J. and Dodd, J.A., \href{https://agupubs.onlinelibrary.wiley.com/doi/full/10.1029/2002JA009688}{Vibrational relaxation of NO (v = 1) by oxygen atoms between 295 and 825 K. Journal of Geophysical Research: Space Physics {\bf{108(A3)}}, 2003.}
\bibitem{mlynzack2021}
Mlynczak, M. G., Hunt, L. A., Lopez$-$Puertas, M., Funke, B., Emmert, J., Solomon, S., ... \& Mertens, C., \href{https://www.sciencedirect.com/science/article/pii/S0022407321001023}{Spectroscopy, gas kinetics, and opacity of thermospheric nitric oxide and implications for analysis of SABER infrared emission measurements at 5.3 $\mu$m. Journal of Quantitative Spectroscopy and Radiative Transfer, 268, 107609, 2021.}
\bibitem{fuller1998}
Fuller‐Rowell, T. J., \href{https://agupubs.onlinelibrary.wiley.com/doi/abs/10.1029/97ja03335}{The “thermospheric spoon”: A mechanism for the semiannual density variation, Journal of Geophysical Research: Space Physics {\bf{103(A3)}}, 2003.}
\bibitem{bharti2018}
Bharti, G., Sunil Krishna, M.V., Bag, T. and Jain, P., \href{https://agupubs.onlinelibrary.wiley.com/doi/full/10.1002/2017JA024576}{Bharti, G., Sunil Krishna, M.V., Bag, T. and Jain, P., 2018. Storm time variation of radiative cooling by nitric oxide as observed by TIMED?SABER and GUVI. Journal of Geophysical Research: Space Physics {\bf{123(2)}}, 2018.}
\bibitem{barth2003}
Barth, C.A., Baker, D.N. and Bailey, S.M., \href{https://agupubs.onlinelibrary.wiley.com/doi/full/10.1029/2003GL018892}{Seasonal variation of auroral electron precipitation. Geophysical research letters {\bf{31(4)}}, 2003.}
\bibitem{duff2003}
Duff, J.W., Dothe, H. and Sharma, R.D., \href{https://agupubs.onlinelibrary.wiley.com/doi/full/10.1029/2002GL016720}{On the rate coefficient of the N($^2$D) + O$_2$ $\rightarrow$ NO + O reaction in the terrestrial thermosphere. Geophysical research letters {\bf{30(5)}}, 2003.}
\bibitem{siskind1992}
Siskind, D.E. and Rusch, D.W., \href{https://agupubs.onlinelibrary.wiley.com/doi/abs/10.1029/91JA02657}{Nitric oxide in the middle to upper thermosphere. Journal of Geophysical Research: Space Physics {\bf{97(A3)}}, 1992.}
\bibitem{swaminathan1998}
Swaminathan, P.K., Strobel, D.F., Kupperman, D.G., Kumar, C.K., Acton, L., DeMajistre, R., Yee, J.H., Paxton, L., Anderson, D.E., Strickland, D.J. and Duff, J.W., \href{https://agupubs.onlinelibrary.wiley.com/doi/abs/10.1029/97JA03249}{Nitric oxide abundance in the mesosphere/lower thermosphere region: Roles of solar soft X rays, suprathermal N($^4$S) atoms, and vertical transport. Journal of Geophysical Research: Space Physics {\bf{103(A6)}}, 1998.}

\end{thebibliography}
\end{document}